\documentclass[showpacs,nofootinbib]{revtex4}
\usepackage[dvips]{graphicx}
\usepackage{amsmath}
\usepackage{mathrsfs}
\usepackage{amsfonts}
\usepackage{amssymb}
\usepackage{bbm}
\usepackage{bm}
\usepackage{braket}

\newcommand{\ud}{\mathrm{d}}

\newcommand{\simorderr}{\raisebox{-4pt}{$\, \stackrel{\textstyle
      <}{\sim} \,$}}

\begin{document}

\title{The influence of strong magnetic fields and instantons on the
  phase structure of the two-flavor NJL model} \author{Jorn
  K. Boomsma} \affiliation{Department of Physics and Astronomy, Vrije
  Universiteit Amsterdam\\ De Boelelaan 1081, NL-1081 HV Amsterdam,
  the Netherlands} \author{Dani\"el Boer} \affiliation{Theory Group,
  KVI, University of Groningen\\ Zernikelaan 25, NL-9747 AA Groningen,
  the Netherlands}

\date{\today}

\begin{abstract}
  Both in heavy-ion collisions as in magnetars very strong magnetic
  fields are produced, which has its influence on the phases of matter
  involved. In this paper we investigate the effect of strong magnetic
  fields ($B \sim 5 m_\pi^2/e = 1.7 \times 10^{19} \rm{G}$) on the
  chiral symmetry restoring phase transition using the
  Nambu-Jona-Lasinio model. It is observed that the pattern of phase
  transitions depends on the relative magnitude of the magnetic field
  and the instanton interaction strength. We study two specific
  regimes in the phase diagram, high chemical potential and zero
  temperature and vice versa, which are of relevance for neutron stars
  and heavy-ion collisions respectively. In order to shed light on the
  behavior of the phase transitions we study the dependence of the
  minima of the effective potential on the occupation of Landau
  levels. We observe a near-degeneracy of multiple minima with
  differing occupation numbers, of which some become the global
  minimum upon changing the magnetic field or the chemical potential.
  These minima differ considerably in the amount of chiral symmetry
  breaking and in some cases also of isospin breaking.
\end{abstract}

%12.39.-x       Phenomenological quark models
%11.30.Rd       Chiral symmetries 
%21.65.Qr       Quark matter 
%25.75.Nq       Quark deconfinement, quark-gluon plasma production and 
%phase transitions in relativistic heavy-ion collisions)

\maketitle

\section{Introduction}
Recently it has been noted that very strong magnetic fields can be
produced in heavy-ion
collisions~\cite{Selyuzhenkov:2005xa,Kharzeev:2004ey,Kharzeev:2007jp}.
Estimates are that at RHIC magnetic fields are created of magnitude
$5.3 m_\pi^2/e = 1.8 \times 10^{19} \rm{G}$ and at LHC of $6 m_\pi^2/e
= 2 \times 10^{19} \rm{G}$, and there are even higher
estimates~\cite{Skokov:2009qp}. Also, certain neutron stars called
magnetars exhibit strong magnetic fields, between $10^{14} - 10^{15}
\rm{G}$~\cite{Duncan:1992hi,Thompson:1993hn}.  These fields occur at
the surface, probably in the much denser interior even higher fields
are present. Using the virial theorem it can be derived that the
maximal strength is $10^{18} - 10^{19} \rm{G}$ \cite{Lai:1991}. If one
assumes that the star is bound by the strong interaction instead of by
gravitation, this limit can be even higher.

In both neutron stars and in heavy-ion collisions it is expected that
quark matter plays a role. Therefore it is interesting to study how
this form of matter behaves in a strong magnetic field. Two different
regions in the QCD phase diagram are of relevance here. Heavy-ion
collisions probe the low chemical potential and high temperature
regime, for neutron stars it is the other way around. In this paper
the effect of very strong magnetic fields will be investigated in both
regimes.

Much work has been done on how an external magnetic field changes
nuclear matter, for a review see Ref.~\cite{Lattimer:2006xb}. The
behavior of ordinary quark matter has been studied using the
Nambu-Jona-Lasinio (NJL) model, see for example
Refs.~\cite{Klevansky:1992qe,Gusynin:1994re,Gusynin:1994va,
  Gusynin:1994xp,Gusynin:1995nb,Ebert:1999ht, Ebert:2003yk,
  Klimenko:2008mg,Inagaki:2003yi,Menezes:2008qt} and recently also in
the linear sigma model coupled to quarks \cite{Fraga:2008qn}. Most
studies investigate the one and two flavor cases, but recently also
the three-flavor case has been investigated
\cite{Osipov:2007je,Menezes:2009uc}.  At high quark chemical
potential, it is believed that the ground state is a color
superconducting phase. The effects of an external magnetic field on
such a phase are discussed in
Refs.~\cite{Ferrer:2005vd,Ferrer:2006vw,Ferrer:2006ie,
  Ferrer:2007uw,Ferrer:2007iw,Noronha:2007wg,Fukushima:2007fc}. Here
color superconductivity will not be considered.

In this paper we study the chiral symmetry restoring phase transition,
which is strongly influenced by an external magnetic field.  From
studies in the NJL model it is known that a magnetic field enhances
the chiral symmetry breaking~\cite{Klevansky:1992qe}, this is related
to the phenomenon of magnetic catalysis of chiral symmetry breaking,
introduced in
Refs.~\cite{Klimenko:1990rh,Klimenko:1992ch,Klimenko:1991he}, further
studied for the NJL model in
Refs.~\cite{Gusynin:1994re,Gusynin:1994va,Gusynin:1994xp,
  Gusynin:1995nb,Ebert:1999ht,Ebert:2003yk,Klimenko:2008mg} and for
QED in e.g.\ Refs.~\cite{Gusynin:1995gt,Lee:1997uh,Lee:1997zj,
  Ferrer:2008dy,Ferrer:2009nq}, where also the generation of an
anomalous magnetic moment was pointed out
\cite{Ferrer:2008dy,Ferrer:2009nq}. This enhancement of chiral
symmetry breaking can be understood as follows, the B-field
anti-aligns the helicities of the quarks and antiquarks, which are
then more strongly bound by the NJL-interaction
\cite{Klevansky:1992qe}. The phenomenon of magnetic catalysis of
chiral symmetry breaking leads to interesting behavior, since it
allows for phases with broken chiral symmetry and nonzero nuclear
density for a range of chemical potentials and magnetic
fields~\cite{Ebert:1999ht,Ebert:2003yk,Inagaki:2003yi}. In such a
phase nonperiodic magnetic oscillations occur, which means that the
constituent quark masses are strongly dependent on the magnetic field,
and consequently also other thermodynamic parameters.

In all studies of the influence of magnetic fields on chiral symmetry
breaking up to now, the effects of instantons have not been studied
explicitly, i.e.\ as a function of instanton interaction strength.
Magnetic fields and instantons can lead to combined effects. In
Ref.~\cite{Kharzeev:2007jp} it is shown that variations in topological
charge, which induce variations of net chirality, in a strong magnetic
field gives rise to an electrical current. This effect is known as the
chiral magnetic effect and could perhaps be observed in heavy-ion
collisions. Variations of topological charge can for instance be
created by instantons.

Here a related study will be performed. We will investigate the
combined effect of instantons and a strong magnetic field on quark
matter using the NJL model. In this model instantons induce an extra
interaction, the 't Hooft determinant interaction, which leads to a
mixing between the different quark flavors. Following the analysis of
Refs.~\cite{Boer:2008ct,Frank:2003ve}, the strength of the instanton
interaction is set by the dimensionless parameter $c$. For $c = 0$
there is no contribution and the quarks are fully independent. Because
of the difference in charge of the quarks, the phase transitions are
decoupled. The other extreme case is $c = 1/2$, which is actually the
most studied case. The quarks are then fully mixed, the constituent
quark masses are equal and the phase transitions will always coincide.
In Ref.\ \cite{Ebert:1999ht} this case is studied in the chiral limit.
It is observed that for a range of typical value for the coupling
constant, phases with broken chiral symmetry and nonzero nuclear
density arise.

In general there is a competition between the magnetic field, which
tends to differentiate the constituent quark masses for different
flavors, and the instanton interaction which favors equal constituent
quark masses. In this work this competition is studied. Apart from
studying the ground state as function of magnetic field and chemical
potential for various characteristic values of $c$, we also look at
the local minima of the effective potential and the corresponding
occupation of Landau levels. It is found that in the neighborhood of
the chiral phase transition the phase diagram develops metastable
phases, differing in the number of filled Landau levels. Some of these
local minima become the global one upon increasing the magnetic field
or chemical potential, but not all of them do. These phases can have
rather different values for the constituent quark masses, in other
words, display significantly different amounts of chiral symmetry
breaking. Unlike in the case of $c=1/2$ which is isospin symmetric, in
these phases the values of the two constituent quark masses can be
very distinct, which corresponds to large isospin violation.
Furthermore, we find that for a realistic choice of parameters,
appearance of phases of broken chiral symmetry and nonzero nuclear
density requires not too large instanton interaction strength, i.e.\ 
$c \simorderr 0.1$.

As mentioned, we also investigate the role of nonzero temperature at
zero chemical potential, which is of relevance for heavy-ion
collisions. Without magnetic field the chiral symmetry restoring 
phase transition at finite temperature is a crossover. 
In the linear sigma model coupled
to quarks it has been observed that the magnetic field turns it 
into a first order transition \cite{Fraga:2008qn}. We
will see that this is not the case for the NJL model. 

This paper is organized as follows. First we derive the effective
potential of the NJL model in the mean-field approximation in a 
magnetic background. Then we discuss the phase diagram as a function
of chemical potential, concentrating on the phase with nonzero
nuclear density and chiral symmetry breaking. We continue with
discussing the temperature dependence and end with conclusions.

\section{Effective potential of the NJL model with a magnetic field}
In this section the effective potential of the NJL model in a strong
magnetic background is derived, this analysis is based on
Refs.~\cite{Fraga:2008qn,Menezes:2008qt}. The effective potential
derived in this way is equal to the one in
Refs.~\cite{Ebert:1999ht,Ebert:2003yk,Klimenko:2008mg} using the
Fock-Schwinger proper time method.

First to set our notation, we briefly review the NJL model introduced
in Refs.~\cite{Nambu:1961tp,Nambu:1961fr}. It is a quark model with
four-point interactions, the gluons are `integrated out'.  In this
paper the following form of the NJL model is used, in the notation of
Ref.~\cite{Boer:2008ct}
\begin{equation}
  \mathcal{L}_{\rm NJL} = \bar \psi \left(i \gamma^\mu \partial_\mu - m - \mu \gamma_0 \right) \psi
  + \mathcal{L}_{\bar q q} + \mathcal{L}_\mathrm{det} \label{lagrangian_NJL},
\end{equation}
where $m = m_u = m_d$ is the current quark mass and $\mu = \mu_u =
\mu_d$ is the quark chemical potential. Note that in contrast to
Ref.~\cite{Boer:2008ct}, here the current quark masses and both
chemical potentials are taken equal.  Furthermore,
\begin{equation}
 \mathcal{L}_{\bar q q} = G_1 \left[ (\bar \psi \tau_a \psi)^2 +  
 (\bar \psi \tau_a i \gamma_5 \psi)^2 \right],
\end{equation}
is the attractive part of the $\bar q q$ channel of the Fierz transformed
color current-current interaction~\cite{Buballa:2003qv} and
\begin{eqnarray}
 \mathcal{L}_\mathrm{det} & = & 8 G_2 
\det \left( \bar \psi_R \psi_L \right) + \mathrm{h.c.} \label{det_int},
\end{eqnarray}
is the 't Hooft determinant interaction which describes the effects of
instantons~\cite{'tHooft:1976fv,'tHooft:1986nc}. Note that this
interaction is flavor mixing. In the literature $G_1$ and $G_2$ are
often taken equal, which means that the low energy spectrum consists
of $\sigma$ and $\bm{\pi}$ fields only, but here we will allow them to
be different. We will restrict to the two flavor case, using $\tau_a$
with $a=0,...,3$ as generators of U(2).

The symmetry structure of the NJL model is very similar to that of
QCD. In the absence of quark masses and the instanton interaction,
there is a global SU(3)$_c \times$U(2)$_L \times$U(2)$_R$-symmetry.
The instanton interaction breaks it to SU(3)$_c
\times$SU(2)$_L\times$SU(2)$_R\times$U(1)$_B$. For nonzero, but equal
quark masses this symmetry is reduced to SU(3)$_c
\times$SU(2)$_V\times$U(1)$_B$. If a magnetic field is turned on, the
symmetry is reduced to SU(3)$_c\times$U(1)$^2$ due to the differences
in charge.

We choose the parameters as in Refs.~\cite{Boer:2008ct,Frank:2003ve}.
This means we write
\begin{equation}
 G_1 = (1 - c) G_0, \quad G_2 = c G_0, \label{values_G1_G2}
\end{equation}
where the parameter $c$ controls the instanton interaction strength,
while the value for the quark condensate (which is determined by the
combination $G_1+G_2$) is kept fixed. For our numerical studies we
will use the following values for the parameters: $m = 6$ MeV, a
three-dimensional momentum UV cut-off $\Lambda = 590$ MeV/$c$ and $G_0
\Lambda^2 = 2.435$. These values lead to a pion mass of $140.2$ MeV, a
pion decay constant of $92.6$ MeV and finally, a quark condensate
$\braket{\bar u u} = \braket{\bar d d} = (-241.5\ \mathrm{MeV})^3$
\cite{Frank:2003ve}, all in reasonable agreement with experimental
determinations.

To calculate the ground state of the theory, the effective potential
has to be minimized. In this section the effective potential is
calculated in the mean-field approximation. We will assume that only
the charge neutral condensates $\braket{\bar \psi \tau_0 \psi}$ and
$\braket{\bar \psi \tau_3 \psi}$ can become nonzero.

To obtain the effective potential in the mean-field approximation,
first the interaction terms are ``linearized'' in the presence of the
$\braket{\bar \psi \tau_0 \psi}$ and $\braket{\bar \psi \tau_3 \psi}$
condensates (this is equivalent to the procedure with a
Hubbard-Stratonovich transformation used in Ref.~\cite{Boer:2008ct})
\begin{equation}
  (\bar \psi \tau_a \psi)^2 \simeq 2 \braket{\bar \psi \tau_a \psi} \bar \psi \tau_a \psi 
  - \braket{\bar \psi \tau_a \psi}^2,
\end{equation}
leading to
\begin{equation}
  \mathcal{L}_{\rm NJL} = \bar \psi \left(i \gamma^\mu \partial_\mu -
  \mathcal{M} - \mu \gamma_0 \right)\psi - \frac{(M_0-m)^2}{4 G_0} -
  \frac{M_3^2}{4 (1-2c)G_0},
\end{equation}
where $\mathcal{M} = M_0 \tau_0 + M_3 \tau_3$ and
\begin{eqnarray}
  M_0 &  =& m - 2 G_0\braket{\bar \psi \tau_0 \psi}, \nonumber \\
  M_3 & = & - 2 (1-2c)G_0\braket{\bar \psi \tau_3 \psi}.
\end{eqnarray}

Now the Lagrangian is quadratic in the quark fields, so we can
integrate over them. After going to imaginary time, this results in
the following effective potential in the mean-field
approximation~\cite{Warringa:2005jh}
\begin{equation}
  \mathcal{V} = \frac{(M_0-m)^2}{4 G_0} + \frac{M_3^2}{4 (1-2c)G_0} - T N_c \sum_{f=u}^d \, \sum_{p_0 = (2 n+1) \pi T} \int \frac{\ud^3 p}{(2 \pi)^3}
  \ln \det \left[ i \gamma_0 p_0 +\gamma_i p_i - M_f - \gamma_0 \mu \right],
\end{equation}
where we have introduced the constituent quark masses $M_u = M_0 +
M_3$ and $M_d = M_0 - M_3$.

As we have mentioned earlier, often $G_1$ is taken equal to $G_2$,
which is the case $c = 1/2$. This choice implies that $M_3$ is then
always equal to 0, i.e., the constituent quark masses are equal. Note
that the reverse is not true.  If the assumption is made that the
constituent quark masses are equal, $c=1/2$ or $\braket{\bar \psi
  \tau_3 \psi}=0$ or both. However, for $M_3=0$ changes in $c$ cannot
be noticed because only the combination $G_1+G_2$ occurs. Hence, the
conclusion is that $\braket{\bar \psi \tau_3 \psi}=0$ and $c$ can be
any value. Instanton effect are nevertheless present ($G_2$ can be
nonzero after all) and $\eta$ and $a_0$ mesons can still be present in
the spectrum.  The ratio $G_2/G_1$ can simply not be determined if
$M_3=0$.  Since magnetic fields affect the two flavors differently
because of the difference in charge, isospin breaking effects are
expected and it is unnatural to choose $M_3=0$ from the start or to
take $c=1/2$. Using the strange quark condensate it was argued in
Ref.~\cite{Frank:2003ve} that a realistic value of $c$ would be around
$0.2$.

\subsection{Including a magnetic field}
Now we include a magnetic field, which changes the dispersion relation
for the quarks in the following way
\begin{equation}
 p^2_{0n} = p_z^2 + M^2 + (2n+1-\sigma)|q|B,
\end{equation}
where $n$ is the quantum number labelling the discrete orbits,
$\sigma$ the spin of the quark, and $q$ is its charge. The integral
over the three-momentum is modified as
\begin{equation}
  \int \frac{\ud^3 p}{(2 \pi)^3} \to \frac{|q|B}{2\pi} \sum_{n=0}^\infty
  \int \frac{\ud p_z}{2 \pi}.
\end{equation}
Performing the sum over the Matsubara frequencies gives the following
effective potential~\cite{Ebert:2003yk,Ebert:1999ht,Menezes:2008qt}
\begin{eqnarray}
 \mathcal{V}& = & \frac{(M_0-m)^2}{4 G_0}+ \frac{M_3^2}{4 (1-2c)G_0} - \frac{N_c}{2 \pi} \sum_{\sigma,n,f} |q_f| B \int
\frac{\ud p_z}{2 \pi} E_{p,f}(B) - \frac{N_c}{2 \pi} \sum_{\sigma,n,f} |q_f|B
\int \frac{\ud p_z}{2 \pi} \Big\{ T \ln \left[ 1+e^{-\left[E_{p,f}(B) + \mu \right] /T} \right] \nonumber \\
&& + T \ln \left[ 1+e^{-\left[E_{p,f}(B) - \mu \right] /T} \right] \Big\} ,
\end{eqnarray}
where $E_{p,f}(B) = \sqrt{p_z^2 + (2 n
  +1-\sigma)|q_f|B+M_f^2}$. Following the analysis of
Ref.~\cite{Menezes:2008qt} this potential can be split in three
pieces, a part that is independent of external parameters, a part that
only depends on the magnetic field and a part that depends on the
magnetic field, chemical potential and temperature
\begin{equation}
 \mathcal{V} = \mathcal{V}_0 + \mathcal{V}_{1}(B) + \mathcal{V}_{2}(B,\mu,T),
\end{equation}
with
\begin{eqnarray}
 \mathcal{V}_0 & = & \frac{(M_0-m)^2}{4 G_0}+ \frac{M_3^2}{4 (1-2c)G_0} - 2 N_c \sum_{f=u}^{d} \int \frac{\ud^3p}{(2\pi)^3}
        \sqrt{\mathbf{p}^2 + M_f^2}, \nonumber \\
 \mathcal{V}_1(B) & = &-\frac{N_c}{2 \pi^2} \sum_{f=u}^{d} (|q_f|B)^2\left[\zeta'(-1,x_f) -
        \frac{1}{2} (x_f^2 -x_f)\ln x_f+\frac{x_f^2}{4}\right],\nonumber \\
 \mathcal{V}_2(B, \mu, T) & = & - \frac{N_c}{2 \pi} \sum_{\sigma,n,f} |q_f|B
\int \frac{\ud p_z}{2 \pi} \Big\{ T \ln \left[ 1+e^{-\left[E_{p,f}(B) + \mu \right] /T} \right]\nonumber \\
&&+ T \ln \left[ 1+e^{-\left[E_{p,f}(B) - \mu \right] /T} \right] \Big\},
\end{eqnarray}
where we have defined $x_f = \frac{M_f^2}{2 |q_f| B}$ and
$\zeta'(-1,x_f) = \frac{d\zeta(z,x_f)}{dz}|_{z=-1}$ with
$\zeta(z,x_f)$ the Hurwitz zeta function. We have neglected $x_f$
  independent terms in $\mathcal{V}_1 (B)$ (including a UV divergent
  one).

The term $\mathcal{V}_0$ is divergent and needs to be regularized.
Here a conventional three-momentum UV cut-off is used, yielding the
expression
\begin{equation}
  \mathcal{V}_0 = \frac{(M_0-m)^2}{4 G_0}+ \frac{M_3^2}{4 (1-2c)G_0} - \frac{N_c}{8 \pi ^2} \sum_{f=u}^{d} |M_f| \left(M_f^3 \ln
      \left(\frac{\Lambda }{M_f} + \sqrt{1+\frac{\Lambda^2}{M_f^2}} \right)-\Lambda 
      \left(M_f^2+2 \Lambda ^2\right) \sqrt{\frac{\Lambda
      ^2}{M_f^2}+1} \,\right).
\end{equation}

The expression $\zeta'(-1,x_f)$ in $\mathcal{V}_1(B)$ can be
written in a more convenient form by differentiating and integrating
the function with respect to $x_f$:
\begin{equation}
  \zeta'(-1,x_f) = \zeta'(-1,0)+\frac{x_f^2}{2}-\frac{x_f}{2}-\frac{x_f}{2} \log (2 \pi)+\psi^{(-2)} (x_f),
\end{equation}
where $\psi^{(m)} (x_f)$ is the $m$-th polygamma function. The term
$\zeta'(-1,0)$ is independent of $x_f$ and will therefore not be taken
into account. The remaining expression is amenable to numerical
evaluation. 

The summation over $\sigma$ and $n$ in $\mathcal{V}_2(B,\mu,T)$ 
can be rewritten as
\begin{equation}
  \mathcal{V}_2(B, \mu, T) = - \frac{N_c}{2 \pi} \sum_{k,f} (2 -
  \delta_{k0} ) |q_f|B
        \int \frac{\ud p_z}{2 \pi} \left\{ T \ln \left[ 1+e^{-\left[ E_{p,k}(T) + \mu \right] /T} \right]
+T \ln \left[ 1+e^{-\left[ E_{p,k}(T) - \mu \right] /T} \right] \right\} , \label{V_medium}
\end{equation}
where $E_{p,k} = \sqrt{p_z^2 + M_f^2 + 2 k |q_f|B}$ and $k$ denotes
the Landau level, which has degeneracy $(2 - \delta_{k0} )$.

At zero temperature, $\mathcal{V}_2$ can be simplified to
\begin{eqnarray}
  \mathcal{V}_2(B, \mu,0) & = & - \frac{N_c}{2 \pi} \sum_{k,f} (2 - \delta_{k0}) \int \frac{\ud p_z}{2 \pi}
    \theta(\mu - E_{p,k}) \left[ \mu - E_{p,k} \right] \nonumber \\
  & = & \sum_{f=u}^d \sum_{k=0}^{k_{f,\mathrm{max}}} \left(2 - \delta_{k0}\right) 
\theta \left(\mu - s_f (k,B) \right) \frac{|q_f| B N_c}{4 \pi^2} \nonumber \\
&& \times \left\{ \mu \sqrt{\mu^2 - s_f^2(k,B)} - s_f^2(k,B) \ln
\left[ \frac{\mu + \sqrt{\mu^2 -s_f^2(k,B)}}{s_f(k,B)} \right] \right\},
\end{eqnarray}
where $s_f(k,B) = \sqrt{M_f^2 + 2 |q_f| B k}$ and $k_{f,\mathrm{max}}$
is the upper Landau level, defined as 
\begin{equation}
  k_{f,\mathrm{max}} = \left\lfloor \frac{\mu^2 - M_f^2}{2 |q_f| B} \right\rfloor,
\end{equation}
where the brackets indicate the floor of the enclosed quantity. 

We will now use these expressions in a numerical study of the minima
of the effective potential, performed along the lines discussed in
Refs.~\cite{Boer:2008ct,Warringa:2005jh}.

\section{Results}

\begin{figure}[htb]
  \includegraphics[scale=0.7]{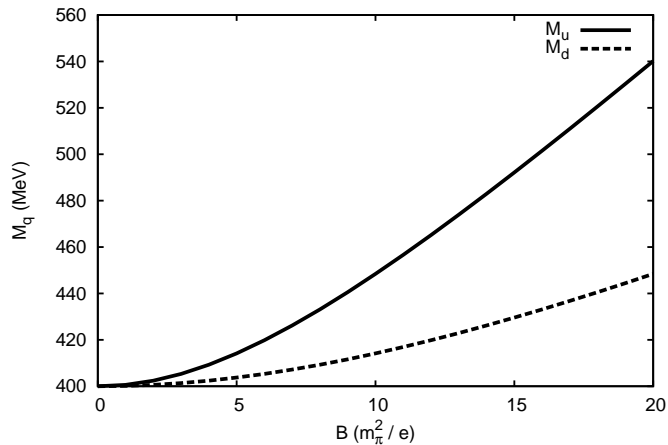}
  \caption{The dependence of the constituent quark masses $M_u$ and
    $M_d$ on the magnetic field $B$.}
  \label{M_vs_B_T0}
\end{figure}

We start with considering the case of $\mu=0, T=0$, and $c=0$.
Fig.~\ref{M_vs_B_T0} shows the results for this unmixed case.  The
magnetic field enhances $M_u$ and $M_d$, which are proportional to
$\braket{\bar u u}$ and $\braket{\bar d d}$, respectively,
consequently the chiral symmetry breaking is enhanced
\cite{Klevansky:1992qe}. Because of the charge difference of the
quarks, the $B$-dependence of the constituent quark masses is not
equal. Nonzero $c$ will cause mixing and will bring the masses closer
to each other.  As discussed, at $c = 1/2$ the constituent quark
masses are exactly equal.

\subsection{Nonzero chemical potential}
In this section we turn to the phase structure near the phase
transition at nonzero chemical potential and zero temperature. From
Refs.~\cite{Klein:2003fy,Toublan:2003tt,Barducci:2004tt,Barducci:2003un}
it is known that when the isospin chemical potential is nonzero, it is
possible to have two phase transitions at low temperature and high
baryon chemical potential. Here we study a similar case, instead of
nonzero isospin chemical potential, we allow for nonzero magnetic
field, we will see that also here the possibility of separate phase
transitions for the two quarks arises. We will take equal chemical
potentials for the quarks, but the magnetic field acts effectively
like a nonzero isospin chemical potential due to the difference in
charge of the quarks.
Instantons cause mixing between the quarks, if the mixing is strong
enough, the two separate phase transitions merge into one. This was
extensively investigated in Ref.~\cite{Klein:2003fy} for the nonzero
isospin chemical potential case.

\mbox{From} Ref.~\cite{Ebert:1999ht,Ebert:2003yk,Inagaki:2003yi},
where the NJL model in the chiral limit was studied, it is known that
Landau quantization induces a more complex phase structure. Apart from
the usual phase of broken chiral symmetry with zero nuclear density,
there is also the possibility of such a phase with nonzero nuclear
density.  Here we perform a more detailed study of this case, which is
a characteristic phenomenon at nonzero chemical potential and
sufficiently strong magnetic fields (cf. Eq.~\eqref{filling} below).

\subsubsection{The $c = 0$ case}
\begin{figure}[htb]
  \includegraphics[scale=.9]{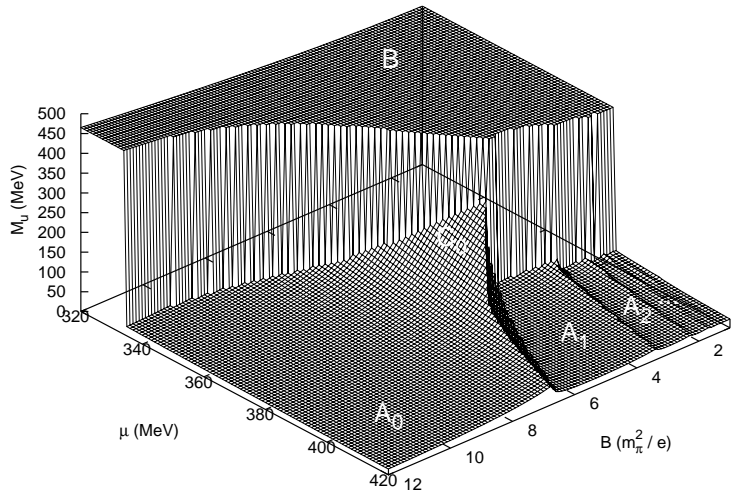}
  \caption{The dependence of the constituent up-quark mass on $B$ and
    $\mu$. $A_i$, $B$ and $C_0$ denote the different phases using the
    scheme of Refs.~\cite{Ebert:1999ht,Ebert:2003yk}. Chiral symmetry
    is broken in phases $B$ and $C_0$, phases $A_0$ and $C_0$ have
    nonzero nuclear density.}
  \label{Mu_vs_B_mu_c0_T0_RS}
\end{figure}

\begin{figure}[htb]
  \includegraphics[scale=.9]{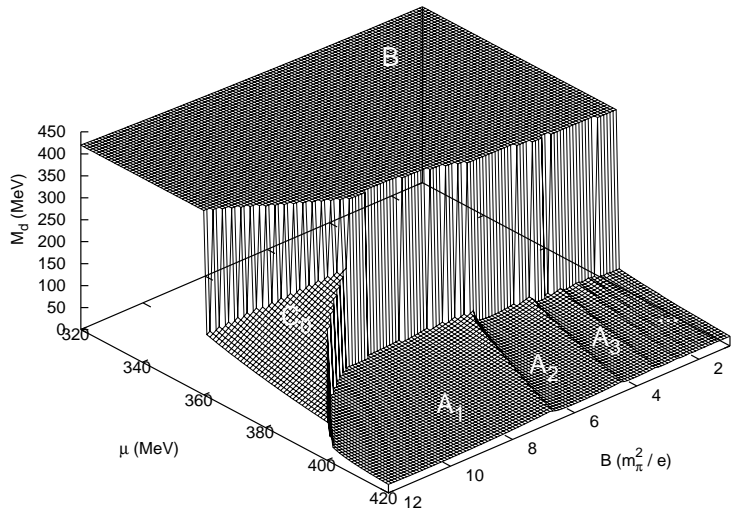}
  \caption{Same as Fig.~\ref{Mu_vs_B_mu_c0_T0_RS}, now for the down
    quark.}
  \label{Md_vs_B_mu_c0_T0_RS}
\end{figure}

When the determinant interaction is turned off, the up and down quarks
are decoupled. This leads to the possibility of separate phase
transitions for the quarks. In Figs.~\ref{Mu_vs_B_mu_c0_T0_RS} and
\ref{Md_vs_B_mu_c0_T0_RS} we show the constituent quark mass of the up
and down quark respectively as a function of quark chemical potential
and magnetic field. As expected, the two quarks have decoupled
behavior.

Let us first discuss the behavior of the up quark. At low chemical
potential we have the ``standard'' chiral symmetry breaking NJL ground
state with empty Landau levels (LL). Following the nomenclature of
Refs.~\cite{Ebert:1999ht,Ebert:2003yk} where the $c=1/2$ case was
studied in the chiral limit, this is called phase $B$.  Note that this
phase always has zero nuclear density. At high chemical potential
chiral symmetry is restored, up to the explicit breaking.  In this
approximate symmetric phase magnetic oscillations can be seen in the
constituent quark masses, caused by Landau quantization. These
oscillating phases are denoted by $A_i$, where $i$ gives the number of
filled LL. As these phases have occupied LL, they have nonzero nuclear
density. The nuclear density of level $k$ is given by
\cite{Menezes:2008qt}
\begin{equation}
  \rho_{f,k} (B, \mu) = (2 - \delta_{k0}) \theta(\mu - s_f(k,B))
  \frac{|q_f| B N_c}{6 \pi^2} \sqrt{\mu^2 - s_f^2
    (k,B)}.\label{filling}
\end{equation}
In the chiral limit the constituent quark masses vanish in the $A_i$
phases.

The oscillations are due to the de Haas-van Alphen effect, which in
QED and in the two-flavor NJL model for $c=1/2$ in the chiral limit
lead to second order transitions between the $A_i$ phases
\cite{Ebert:1999ht}. However, with our choice of parameters the
transitions are weakly first order. In the chiral limit they become
second order, like for $c=1/2$, as can for instance be seen in the
nuclear density. For completeness, we mention that in the color
superconducting case of Ref.~\cite{Noronha:2007wg,Fukushima:2007fc}
the oscillations in the gap parameter are seen to be continuous, but
second order transitions can occur when neutrality conditions are
imposed.

For $B$ larger than $4.5 m_\pi^2/e$ an interesting intermediate phase
arises, where the up-quark jumps as a function of $\mu$ first to a
phase with a still rather large constituent mass and then to phase
$A_1$. This intermediate phase is called $C_0$ in the language of
Refs.~\cite{Ebert:1999ht,Ebert:2003yk} and corresponds to a phase of
broken chiral symmetry having nonzero nuclear density and a filled
zeroth LL. So the essential difference between this phase and
$A_0$ is the breaking of chiral symmetry. For smaller values of the
coupling constant $G_0$ also the phases $C_k$ with $k > 0$ (which are
similar to $C_0$ but with more occupied LL) occur. The transitions
between the $C_k$ are first order, furthermore, they are nonperiodic
in the sense that the difference between the transitions is
$B$-dependent as the constituent mass strongly depends on $B$
\cite{Ebert:1999ht}. If we are in this phase $C_0$ and increase the
magnetic field, the constituent quark mass decreases, eventually
becoming almost zero, this can be interpreted as a crossover to $A_0$.
In the chiral limit the crossover becomes a second order transition.
Finally, we would like to note that already at $B = 4 m_\pi^2/ e$ the
phase $C_0$ exists as a metastable phase (we will discuss this in more
detail later).

The qualitative behavior of the down quark is very similar, as the
quarks only differ in charge, consequently
Fig.~\ref{Md_vs_B_mu_c0_T0_RS} can be directly obtained from
Fig.~\ref{Mu_vs_B_mu_c0_T0_RS} by multiplying $B$ by 2, for ease of
comparison we show both figures. If one compares the two figures, one
can immediately see that there are large regions where the constituent
quark masses are considerably different. This is equivalent to a large
nonzero $\braket{\bar \psi \tau_3 \psi}$ condensate, i.e., spontaneous
isospin breaking. This will influence the behavior of the mesons
accordingly, for example the masses.

Eventually, if one keeps increasing the magnetic field, the phase
transitions of the quarks will take place at (almost) the same
chemical potential and there will be no spontaneous isospin breaking.

\subsubsection{The $c \neq 0$ case}
In this section the consequences of the instanton
interaction is studied, i.e., the parameter $c$ is varied. Increasing
$c$ will cause mixing between the constituent quarks, which tends to
bring the constituent quark masses together. Around the phase
transition there is a competition between the effect of the magnetic
field and the instanton interaction.

\begin{figure}[htb]
  \includegraphics[width=\textwidth]{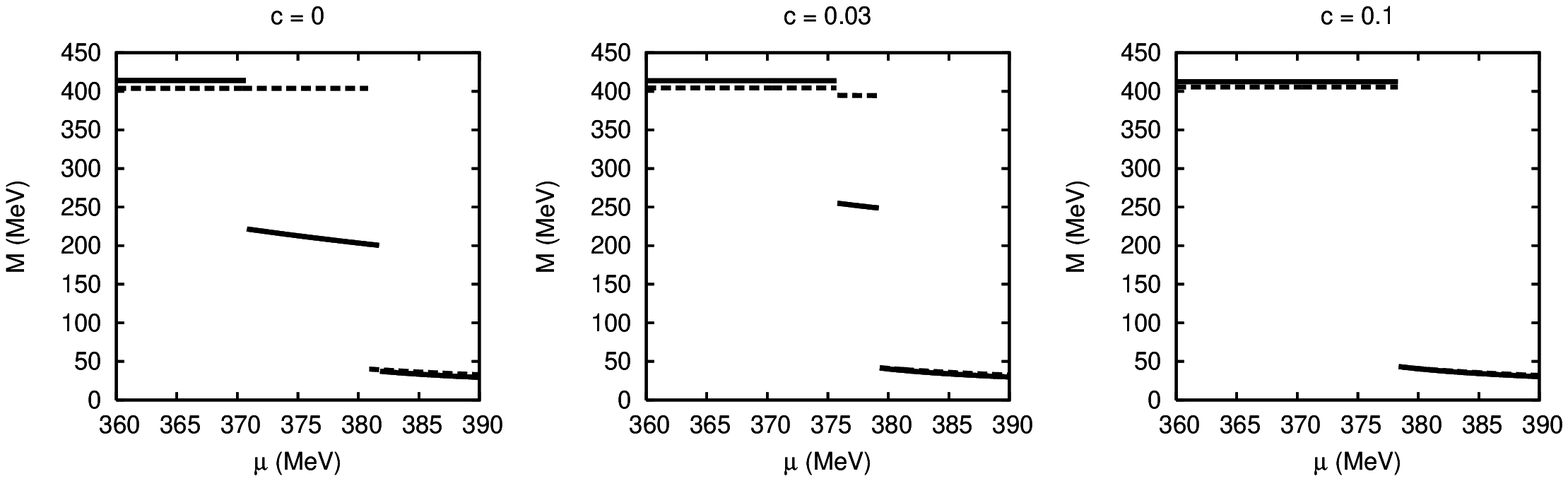}
  \caption{The dependence of the constituent quark masses on the quark
    chemical potential for $B = 5 m_\pi^2/e$ and various $c$ values. Solid
    lines denote the up quarks, the dashed lines the down quark.}
  \label{M_vs_mu_T0_B5mpi2}
\end{figure}

The competition is illustrated in Fig.~\ref{M_vs_mu_T0_B5mpi2}, where
the constituent quark masses are plotted as a function of the quark
chemical potential for three characteristic values for $c$, $c = 0$,
$0.03$, and $0.1$ with $B = 5 m_\pi^2/e$. The qualitative behavior for
different values of the magnetic field is similar.  One can see that
when $c \neq 0 $, the phase transitions are indeed coupled.
Furthermore, one observes that the two phase transitions merge into
one when $c$ is increased and that the phase $C_0$ disappears.
Qualitatively the behavior is similar to the case of nonzero isospin
chemical potential studied in Ref.~\cite{Klein:2003fy}, but in that
case the phase $C_0$ does not exist.

When the coupling constant $G_0$ is lowered, it is possible to have
$C_k$ phases at $c = 1/2$, as in Ref.\ \cite{Ebert:1999ht}.  Compared
to the chiral limit studied there, the region of the phase diagram
with $C_k$ phases increases for $m\neq 0$.

\begin{figure}[b]
  \includegraphics[scale=0.6]{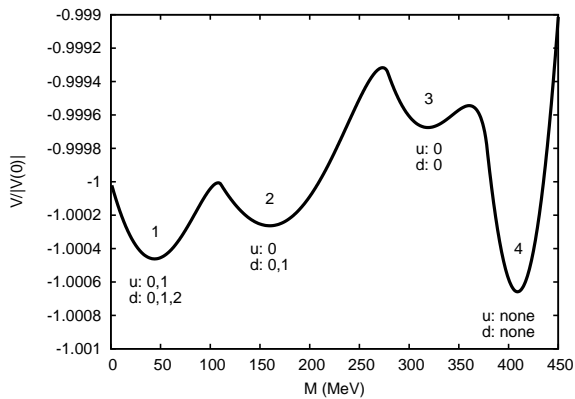}
  \caption{The normalized effective potential at the values $B = 5
    m_\pi^2/e$, $\mu = 378 \, \rm{MeV}$ and $c = 1/2$.  There are four
    minima. The numbers below the minima denote the LL occupied for
    each quark. Note that the minima are almost degenerate.}
  \label{effpot_B5mpi2_c0.5_mu178}
\end{figure}

More insight into the phase structure and phase transitions is
obtained by looking at the behavior of local minima of the effective
potential. Near the phase transition at these (large) magnetic fields,
metastable phases arise. These phases differ in the number of filled
LL. Let us take as an example the $c = 1/2$ case, which is the easiest
to discuss, as the effective potential is then only a function of $M_u
= M_d = M$. In Fig.~\ref{effpot_B5mpi2_c0.5_mu178} we show the
effective potential as a function of $M$ with $\mu = 378 \, \rm{MeV}$
and $B = 5 m_\pi^2/e$.  At these values four minima can be seen, the
global minimum is the phase in which the chiral symmetry breaking is
largest, i.e.\ minimum $4$. When $\mu$ is increased, minimum $1$ will
take over, which is $A_1$ for the up quark and $A_2$ for the down
quark. The other two local minima never become the global one for our
choice of $G_0$, but as they are almost degenerate with the other
minima (also for other values of $c$), they are nevertheless
important. These local minima correspond to $C_k$ phases and can
become the global minimum when $G_0$ is lowered.

Similar results hold for $c \neq 1/2$, also then metastable phases
exists with different fillings of LL. In this cases some of the $C_k$
phases can become the global minimum, as we have seen for $c=0$. Like
before, the number of such states depends on the choice of the other
parameters.

As the metastable phases differ in the values of the $\braket{\bar u
  u}$ and $\braket{\bar d d}$ condensates at small $c$, they again
represent rather large broken isospin and will lead to different meson
masses. Whenever the system is passing the phase transition, it could
be trapped in one of those metastable phases for some time and
consequences from the changing meson masses can arise, for example,
enhancing or suppressing certain decays.

\subsection{Nonzero temperature}
In this section the temperature dependence of the ground state is
investigated at zero chemical potential, but with a magnetic field.
As the instanton interaction does not influence
the temperature dependence much, we only consider $c = 1/2$ for
simplicity. Ref.~\cite{Fraga:2008qn} found in the
linear sigma model coupled to quarks, the usual crossover becomes a
first order transition at very high magnetic fields. However, we find that
this is not the case in the NJL model.

In Ref.~\cite{Fraga:2008qn} only the lowest Landau level was taken
into account. Here more Landau levels are included, so the effect of
the higher Landau levels can be investigated in the NJL model.  Since
the levels with large $k$ are exponentially suppressed, the summation
can be truncated in Eq.~\eqref{V_medium}, we will denote the largest
$k$ with $k_\mathrm{trunc}$.  The value of $k_\mathrm{trunc}$ depends
on the temperature, constituent quark mass, chemical potential and
magnetic field considered. If $M$ and $T$ are increased or if $B$ is
decreased, $k_\mathrm{trunc}$ has to be increased.

\begin{figure}[htb]
  \includegraphics[scale=0.6]{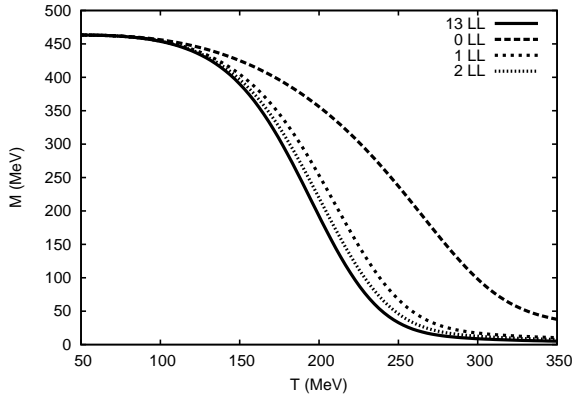}
  \caption{The temperature dependence of the constituent quark mass
    for strong magnetic field ($B = 15 m_\pi^2/e$) and various
    $k_{\rm{trunc}}$ values.}
  \label{M_vs_T_B_15mpi2_LL}
\end{figure}

In Fig.~\ref{M_vs_T_B_15mpi2_LL} we show the temperature dependence of
the constituent quark mass at $B = 15 m_\pi^2/e$ for four different
values of $k_\mathrm{trunc}$.  The 13 levels case is chosen such that
the error is less than 1 percent at $M = 450 \, \rm{MeV}, T = 450 \,
\rm{MeV}$. From the figure it can be inferred that taking more Landau
levels into account, makes the crossover sharper. Also, there is a
significant difference between including the zeroth and first Landau
level. It is clear that including more Landau levels, influences the
details of the transition.  However, the qualitative aspects of the
phase transition are not changed.

\begin{figure}[htb]
  \includegraphics[scale=0.6]{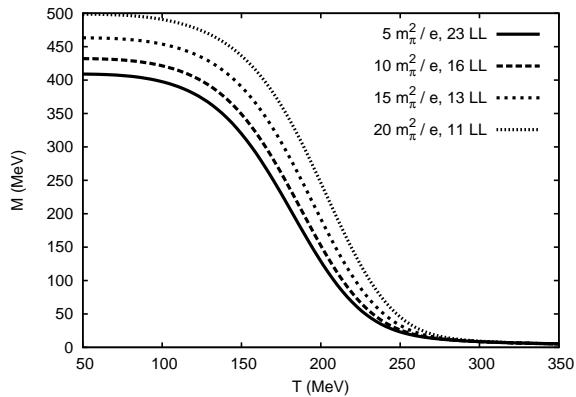}
  \caption{The temperature dependence of the constituent quark mass
    for various strong magnetic fields. We also indicate $k_{\rm{trunc}}$.}
  \label{M_vs_T_B}
\end{figure}

In Fig.~\ref{M_vs_T_B} the temperature dependence of the constituent
quark mass for different values of the external magnetic field is
shown. The phase transition remains a crossover, in contrast to the
results in the linear sigma model coupled to quarks. This difference
is important, as a first order phase transition allows for meta-stable
states, whereas a crossover does not.

\begin{figure}[htb]
  \includegraphics[scale=0.6]{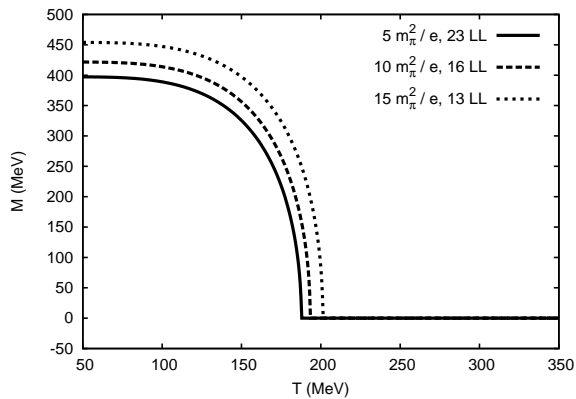}
  \caption{Same as Fig.~\ref{M_vs_T_B}, now in the chiral limit.}
  \label{M_vs_T_B_chiral}
\end{figure}

In Fig.~\ref{M_vs_T_B_chiral} the results in the chiral limit are
shown, where the transition remains a second order phase transition,
like it is the case at zero magnetic field and confirms the results of
\cite{Inagaki:2003yi} who calculated the phase diagram in a strong
magnetic field in the chiral limit using the Fock-Schwinger proper
time method. Note that the critical temperature increases slightly
with increasing magnetic field.

\section{Conclusions} 
The effect of a strong magnetic field on quark matter has been
investigated in the NJL model in two regimes, zero temperature and
finite chemical potential and vice versa. The first regime is of
relevance for (the interior of) magnetars and the second for heavy-ion
collisions.

At very high magnetic fields, when $M \approx 2 |q| B \approx \mu$,
the phase structure shows a variety of phases and phase transitions
due to Landau quantization. As a function of chemical potential, more
phase transitions occur, corresponding to Landau levels filling up
successively. Due to the difference in charge, this pattern is
different for the two quark flavors. When there is no mixing in the
absence of the instanton interaction, the two patterns are uncoupled.
This generally leads to rather different constituent quark masses, or
equivalently, spontaneous isospin breaking $\braket{\bar \psi \tau_3
  \psi} \neq 0$. This affects the mesons inside the medium, for
example their masses. It was found that for a realistic choice of
parameters in the NJL model such a phase of broken chiral and isospin
symmetry arises around $B = 4.5 m_\pi^2/e$, but it is already present
as a metastable phase for lower magnetic fields.

When the instanton interaction is included, a competition occurs
between the strength $c$ of this interaction and the magnetic field.
This reduces the region in the phase diagram with large $\braket{\bar
  \psi \tau_3 \psi}$. For $c$ sufficiently large it disappears
entirely, leaving only one phase transition. However, around this
transition the phase structure is still rather complex regarding
metastable phases, which are characterized by different fillings of
Landau levels and which differ only slightly in energy, but much in
the amount of chiral symmetry breaking. For lower values of $c$ some
of these near-degenerate minima can also differ considerably in the
amount of isospin breaking.

Finally the role of temperature was studied at zero chemical
potential. In Ref.~\cite{Fraga:2008qn} it was found in the linear
sigma model coupled to quarks that a strong magnetic field changes the
usual crossover as a function of temperature into a first order
transition. In the NJL model it was found that the crossover remains
a crossover. Also it was found that including higher Landau levels in
the calculation of the effective potential changes the
details of the crossover, it becomes sharper, albeit the qualitative
aspects of the transition are not changed. The difference between the
two models is important, as the first order transition allows for
metastable phases, while a crossover does not.

\begin{acknowledgments}
  We would like to thank E. Fraga and A. Mizher for useful
  suggestions. Furthermore, we would like to thank A. Mizher for her
  help with implementing the magnetic field.
\end{acknowledgments}

\end{document}